\documentclass{article}
\usepackage{spconf,amsmath,graphicx,hyperref}
\usepackage{booktabs,multirow,makecell}
\usepackage{colortbl,xcolor}
\usepackage{amssymb}
\definecolor{bestrow}{HTML}{E8F5E9}
\definecolor{headerblue}{HTML}{1A237E}
\definecolor{headerbg}{HTML}{E8EAF6}
\newcommand{\ci}[2]{%
  \makecell[c]{%
    #1\\[-2.4pt]
    {\fontsize{5}{5.2}\selectfont #2}%
  }%
}

\title{Bearings: Self-Supervised Soundfield Embeddings from First-Order Ambisonics}

\name{Goksenin Yuksel\textsuperscript{1}, Marcel van Gerven\textsuperscript{1}, Kiki van der Heijden\textsuperscript{1,2}}
\address{\textsuperscript{1}Dept. of Machine Learning and Neural Computing, Donders Institute,\\
Radboud University, Nijmegen, the Netherlands\\
\textsuperscript{2}Mortimer B. Zuckerman Institute, Columbia University, New York, USA}

\begin{document}
\ninept
\maketitle
\begin{abstract}
Recently proposed self-supervised audio encoders learn powerful general-purpose representations of sound scenes, yet they are spatially blind. To supply the missing spatial representation of sound scenes, we introduce \textsc{Bearings}.   \textsc{Bearings} is a self-supervised framework that learns soundfield embeddings from unlabeled first-order Ambisonics. We pre-train a masked auto-encoder paired with a decoder conditioned on frozen acoustic embeddings from an off-the-shelf single-channel audio encoder. Our results show that the resulting soundfield embeddings form a reusable stream that can be attached to frozen acoustic 
encoders with a lightweight trainable fusion head. On sound event localization and detection, concatenating our soundfield embeddings with acoustic representations provides the missing spatial information and enables joint detection and localization, raising the location-dependent F-score from below 4 to 50 on TAU-NIGENS 2021 and 39 on STARSS23. To our knowledge, \textsc{Bearings} is the first self-supervised soundfield encoder whose embeddings plug into frozen acoustic encoders without retraining either model.
\end{abstract}
\section{Introduction}
\label{sec:intro}
Self-supervised general-purpose audio representation learning has seen remarkable progress: models such as ATST~\cite{atst}, BEATs~\cite{beats}, Dasheng~\cite{dasheng}, SPEAR~\cite{spear} learn powerful general-purpose embeddings from large-scale unlabeled audio. Despite their acoustic fidelity, these models are inherently blind to spatial information: they encode only \emph{what} sounds are present, not \emph{where} those sounds originate.

Joint spatial and acoustic perception has traditionally been addressed through supervised sound event localization and detection (SELD)~\cite{seld}, where models such as SELDNet~\cite{seldnet} map multichannel audio features to joint sound detection and direction-of-arrival estimates. More recently, pre-training has been extended to spatial audio: w2v-SELD~\cite{w2v-seld} adapts contrastive waveform pre-training to Ambisonics for SELD; Spatial-AST---the spatial front-end of BAT~\cite{bat}---is trained on synthetic binaural mixtures for joint localization, detection, and LLM-based spatial reasoning; and ELSA~\cite{elsa} aligns spatial audio with text representations for 3D descriptions of the sound scenes. Closest to our setting, GRAM-Ambisonics~\cite{gram} extends self-supervised pre-training to Ambisonics, learning content and spatial structure jointly. These approaches entangle spatial and acoustic modeling within a single backbone~\cite{bat, gram, elsa}. The SELD approaches that were trained on real-life recordings are optimized for a particular SELD setup~\cite{w2v-seld, seldnet}. Concurrently, the emerging class of audio-LLMs~\cite{qwenaudio, salmonn} has advanced high-level acoustic reasoning, yet they remain spatially constrained by their reliance on single-channel encoders, with only limited attempts at spatial grounding~\cite{bat, sciphi}. While such spatial audio-LLMs can reason over static sound scenes, they do not expose spatial representations of the sound scene.

We introduce \textsc{Bearings}, a modular framework that learns reusable soundfield embeddings through spatial self-supervision. Once pre-trained, the soundfield encoder can be paired with different single-channel acoustic encoders while keeping both encoders frozen. This allows the downstream acoustic encoder to be replaced as stronger models become available, without repeating spatial pre-training. 

\begin{figure*}[!t]
    \centering
\includegraphics[width=0.9\linewidth]{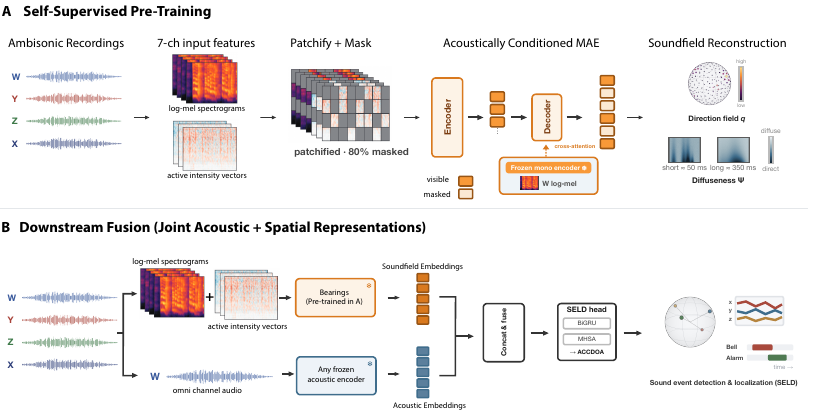}
    \caption{\textsc{Bearings} overview. \textbf{(A)} Self-supervised pre-training: a masked ViT encodes seven-channel FOA features (four log-mel spectrograms, three active-intensity vectors), and a decoder that cross-attends to embeddings from a frozen mono encoder reconstructs direction distributions and two-scale diffuseness at the masked patches. \textbf{(B)} Downstream fusion: the frozen \textsc{Bearings} encoder supplies soundfield embeddings that are concatenated with acoustic embeddings from any frozen single-channel acoustic encoder.}
    \label{fig:overview}
\end{figure*}

We learn soundfield embeddings in a fully self-supervised manner from unlabelled in-the-wild clips crawled from 360$^{\circ}$ YouTube videos~\cite{ytambigen}. From the FOA signal, we extract a 7-channel spatial representation consisting of four log-mel spectrograms and three normalized active-intensity vectors. The soundfield encoder is trained as a masked auto-encoder whose decoder cross-attends to acoustic embeddings from a frozen single-channel encoder while reconstructing the direction field and diffuseness~\cite{dirac} (Fig.~\ref{fig:overview}).

Downstream fusion experiments validate that~\textsc{Bearings}\footnote{Code\&Models: \url{https://github.com/labhamlet/Bearings}} embeddings capture substantial spatial information. Fused with monaural acoustic content embeddings via lightweight heads, they raise the location-dependent F-score from below 4 to 41--50 on TAU-NIGENS 2021, and 34--39 on STARSS23 across three architecturally distinct acoustic backbones. With GRAM-Clean as the content encoder, the fused model surpasses the supervised SELDNet baseline~\cite{seldnet} on every metric of both datasets except $\mathrm{LE}_{\mathrm{CD}}$ on STARSS23, where the two tie at $22.0^\circ$, and surpasses the jointly trained GRAM-Ambisonics~\cite{gram} on TAU-NIGENS 2021 while matching it on STARSS23. These results establish that soundfield representations can be learned without spatial annotations and reused across frozen acoustic encoders.

\section{Methodology}
\label{sec:methodology}
\subsection{Pre-Training Data}
\label{sec:pretraining_data}
We pre-train the soundfield encoder on YT-AmbiGen~\cite{ytambigen}, a dataset containing roughly 102 K five-second audio clips in first-order Ambisonics (FOA) format, which are sourced from $360^{\circ}$ YouTube videos. The FOA signal consists of an omnidirectional channel $B_{00}(t)$ and three directional channels $B_{1,-1}(t)$, $B_{10}(t)$, $B_{11}(t)$. The clips capture real-world sound scenes with dynamic as well as static sound sources, exposing the model to naturalistic soundfields during pre-training.

\subsection{Input Features}
\label{sec:features}
The soundfield encoder takes as input a 7-channel spatial time-frequency representation $\mathbf{F}(t,b) \in \mathbb{R}^{7 \times T \times B}$ extracted from the FOA signal, consisting of four log-mel spectrograms and three normalized active-intensity vectors. First, we normalize the FOA waveforms by  $B_{00}$ loudness. We then compute the log-mel spectrogram per channel (short-time Fourier transform, window size = 1024, hop length = 10 ms, 128 mel bands). Additionally, we compute the Cartesian active
intensity at each time--frequency bin as
\begin{equation}
\mathbf{I}_a(t,f) = 2\,\mathrm{Re}\!\left[\hat{B}_{00}^{*}(t,f)\cdot
\big(\hat{B}_{1,-1}, \hat{B}_{10}, \hat{B}_{11}\big)^{\top}\right],
\end{equation}
with total energy
$E(t,f) = |\hat B_{00}|^2 + |\hat B_{1,-1}|^2 + |\hat B_{10}|^2 + |\hat B_{11}|^2$.
We aggregate intensity and energy to the mel scale separately and
normalize per band, resulting in the active-intensity vector (AIV)
\begin{equation}
\label{eq:aiv}
\hat{\mathbf{I}}_a(t,b) = \frac{\sum_f m_b(f)\,\mathbf{I}_a(t,f)}
{\sum_f m_b(f)\,E(t,f) + \epsilon}.
\end{equation}

\subsection{Soundfield Encoder Architecture}
\label{sec:spatial_encoder}
We tokenize the input with a single non-overlapping 2D convolution that projects each time--frequency patch (shape $f \times t$ with $f$ = 16 and $t$ = 8) across all $C=7$ channels jointly to dimension $D$.
We add fixed 2D sine--cosine positional embeddings and pass the sequence through the soundfield encoder $f_\psi$, which is a shallow standard ViT backbone~\cite{dosovitskiy2020vit} (6 layers, width $D{=}384$, 10.9\,M parameters). The resulting patch embeddings $\hat{\mathbf{z}}_1,\dots,\hat{\mathbf{z}}_P\in\mathbb{R}^D$ serve as
frame-level soundfield embeddings for downstream tasks.

\subsection{Acoustically-Conditioned Masked Reconstruction}
\label{sec:ssl_objectives}
We train $f_\psi$ as a masked auto-encoder (MAE)~\cite{mae}. The encoder operates on a small
set of visible patches while the decoder reconstructs the spatial field at masked locations conditioned on acoustic embeddings extracted from a
single-channel acoustic encoder. 

\noindent\textbf{Masking and encoding.} We mask $80\%$ of the $P$ patches (tube masking
across all seven channels at each masked location) and encode only the visible
patches, yielding $\mathbf{z}_{\mathrm{vis}}\in\mathbb{R}^{P_{\mathrm{vis}}\times
D}$. We use a masking strategy that masks (i) the full time axis of the input features
($40\%$), (ii) random patches ($20\%$), and (iii) right-censored future frames ($40\%$). 

\noindent\textbf{Conditioned decoder.} The encoder outputs and a shared learnable mask token at masked positions are fed into the decoder together with 2D sine--cosine positional embeddings. The decoder is a 4-layer ViT (width $D$ = 256, 8 attention heads, 4.6 M parameters). Each decoder block applies
(i) self-attention over the patch embeddings and (ii) cross-attention into a
single-channel acoustic embedding sequence $\mathbf{C}=g_\theta(\mathrm{Enc}_{\mathrm{mono}}(M_{00}))$
obtained by passing the $B_{00}$ log-mel spectrogram through a frozen GRAM-Clean
encoder (the mono encoder of GRAM)~\cite{gram}.

\noindent\textbf{Soundfield reconstruction.} At each masked patch the decoder predicts two
analytically derived spatial features. 

\noindent\emph{(i) Direction.} We predict a categorical distribution over a
Fibonacci grid $\mathcal{G}$ of $G{=}256$.
The target is an energy-weighted kernel density estimate of the local intensity direction: after smoothing the mel intensity over a small
time--frequency tile (shared with the short diffuseness window below), each
tile contributes a von Mises--Fisher bump centred on its smoothed direction
$u_\tau=\bar{\mathbf{I}}_\tau/\lVert\bar{\mathbf{I}}_\tau\rVert$, weighted by
its smoothed intensity magnitude,
\begin{equation}
q_{p,g}\;\propto\;\sum_{\tau\in p}
\lVert\bar{\mathbf{I}}_\tau\rVert\,
\exp\!\big(\kappa\,(g^{\top}u_\tau-1)\big),\qquad
\kappa=G/2\pi.
\end{equation}
The target is normalized so that $\sum_g q_{p,g}=1$. We supervise it with a weighted
cross-entropy over the masked set,
\begin{equation}
\mathcal{L}_{q}=\frac{\sum_{p\in\mathcal{M}} \omega_p\big(-\sum_g q_{p,g}\log\hat q_{p,g}\big)}
{\sum_{p\in\mathcal{M}} \omega_p},\qquad
\omega_p=\frac{W_p}{W_p+\bar W},
\end{equation}
where $W_p=\sum_{\tau\in p}\lVert\bar{\mathbf{I}}_\tau\rVert$ is the patch's
accumulated directional energy and $\bar W$ the batch mean of $W_p$.

\noindent\emph{(ii) Diffuseness (two time scales).} Following the DirAC~\cite{dirac} formulation,
diffuseness is one minus the ratio of the magnitude of the time-averaged active
intensity to the time-averaged energy,
\begin{equation}
\Psi^{(s)}(t,b)=1-\frac{\big\lVert\langle\mathbf{I}^{\mathrm{mel}}_a\rangle_{\tau_s}\big\rVert}
{c\,\langle E^{\mathrm{mel}}\rangle_{\tau_s}+\epsilon},
\end{equation}
computed from the \emph{raw} mel intensity and energy, with $c=1$ under SN3D
normalization guaranteeing $\Psi\in[0,1]$. We predict two scales, a short window
$\tau_{\mathrm{s}}\!\approx\!50$\,ms and a long window
$\tau_{\mathrm{l}}\!\approx\!350$\,ms. The loss is an $\ell_1$ term,
$\mathcal{L}_\Psi=\frac{1}{N}\sum_i|\hat\Psi_i-\Psi_i|$.
The total objective is
\begin{equation}
\mathcal{L}=\mathcal{L}_q+\mathcal{L}_\Psi.
\end{equation}

\subsection{Downstream SELD Evaluation}
\label{sec:seld_eval}

\begin{table*}[!t]
\centering
\caption{Downstream SELD performance. Bold marks the best value per column.}
\label{tab:downstream}
\setlength{\tabcolsep}{2.2pt}
\renewcommand{\arraystretch}{0.92}
\scriptsize
\begin{tabular}{ll cccc cccc}
\toprule
& &
\multicolumn{4}{c}{\textbf{TAU-NIGENS 2021}} &
\multicolumn{4}{c}{\textbf{STARSS23}} \\
\cmidrule(lr){3-6}
\cmidrule(lr){7-10}
\textbf{Backbone}
& \textbf{Method}
& $\mathrm{ER}_{20^\circ}\downarrow$
& $\mathrm{F}_{20^\circ}\uparrow$
& $\mathrm{LE}_{\mathrm{CD}}\downarrow$
& $\mathrm{LR}_{\mathrm{CD}}\uparrow$
& $\mathrm{ER}_{20^\circ}\downarrow$
& $\mathrm{F}_{20^\circ}\uparrow$
& $\mathrm{LE}_{\mathrm{CD}}\downarrow$
& $\mathrm{LR}_{\mathrm{CD}}\uparrow$ \\
\midrule
& SELDNet (7-ch FOA + IVs)~\cite{dcase2021, starss23}
& 0.69
& 33.9
& 24.1
& 43.9
& 0.57
& 29.9
& 22.0
& 47.7 \\
& GRAM-Ambisonics (7-ch FOA + IVs)~\cite{gram}
& 0.62
& 41.0
& 23.3
& 66.3
& 0.56
& 37.4
& 22.8
& \textbf{58.8} \\
\midrule
\multirow{2}{*}{\makecell[l]{GRAM-Clean}}
& Single-channel, acoustic only
& \ci{0.83}{[0.81,0.84]}
& \ci{2.8}{[2.3,3.3]}
& \ci{102.1}{[99.3,104.9]}
& \ci{34.7}{[33.0,36.2]}
& \ci{0.83}{[0.81,0.86]}
& \ci{2.5}{[1.6,3.4]}
& \ci{147.8}{[146.2,149.3]}
& \ci{16.6}{[15.0,18.2]} \\
& + Bearings (ours)
& \ci{\textbf{0.55}}{[0.53,0.56]}
& \ci{\textbf{50.4}}{[48.2,52.7]}
& \ci{\textbf{19.3}}{[18.3,20.3]}
& \ci{\textbf{66.8}}{[65.0,68.5]}
& \ci{\textbf{0.51}}{[0.46,0.55]}
& \ci{\textbf{38.9}}{[34.6,43.8]}
& \ci{22.0}{[18.3,25.7]}
& \ci{55.5}{[49.3,61.4]} \\
\midrule
\multirow{2}{*}{\makecell[l]{Dasheng-Base}}
& Single-channel, acoustic only
& \ci{0.83}{[0.82,0.85]}
& \ci{3.2}{[2.6,3.9]}
& \ci{87.1}{[84.1,90.3]}
& \ci{43.2}{[41.5,45.0]}
& \ci{0.82}{[0.80,0.86]}
& \ci{2.9}{[2.3,3.5]}
& \ci{146.1}{[144.6,147.5]}
& \ci{18.9}{[17.1,20.7]} \\
& + Bearings (ours)
& \ci{0.56}{[0.55,0.58]}
& \ci{47.7}{[45.7,50.1]}
& \ci{20.0}{[19.0,21.0]}
& \ci{65.0}{[62.9,67.2]}
& \ci{0.54}{[0.50,0.58]}
& \ci{34.3}{[30.2,38.7]}
& \ci{24.9}{[21.5,29.3]}
& \ci{53.8}{[48.7,59.1]} \\
\midrule
\multirow{2}{*}{\makecell[l]{SPEAR-Base}}
& Single-channel, acoustic only
& \ci{0.89}{[0.87,0.90]}
& \ci{3.0}{[2.5,3.5]}
& \ci{86.4}{[83.3,89.8]}
& \ci{36.5}{[34.8,38.4]}
& \ci{0.86}{[0.84,0.89]}
& \ci{2.4}{[1.5,3.3]}
& \ci{145.9}{[142.1,149.9]}
& \ci{11.3}{[9.9,12.8]} \\
& + Bearings (ours)
& \ci{0.61}{[0.59,0.64]}
& \ci{40.9}{[38.3,43.5]}
& \ci{22.2}{[20.9,23.4]}
& \ci{61.5}{[59.1,63.8]}
& \ci{0.56}{[0.53,0.60]}
& \ci{34.9}{[30.5,39.0]}
& \ci{\textbf{21.2}}{[18.0,24.3]}
& \ci{49.0}{[43.7,53.6]} \\
\bottomrule
\end{tabular}
\end{table*}

We attach the frozen soundfield embeddings to frozen, state-of-the-art monaural acoustic encoders and train a lightweight head to perform SELD, a task that inherently requires joint reasoning over acoustic content and
spatial directionality~\cite{seld}. Here, the fusion combines acoustic embeddings with the spatial information captured by \textsc{Bearings}. We evaluate in two complementary settings: scenes generated with measured room impulse responses (TAU-NIGENS 2021~\cite{dcase2021}) and real recordings captured
with an Ambisonic microphone array (STARSS23~\cite{starss23}). For STARSS23, we use the synthetic + real dataset as in the DCASE 2023 Challenge 3 baseline. 

To test versatility, we pair the frozen soundfield encoder with three architecturally distinct frozen backbones, all of which process only the $B_{00}$ channel: GRAM-Clean~\cite{gram} (time--frequency patch
tokens), Dasheng-Base~\cite{dasheng} (time-only tokens), and SPEAR~\cite{spear} (waveform). Tokens with an explicit frequency axis (\textsc{Bearings}, GRAM-Clean) are first pooled over frequency using learned frequency attention. All sequences are then temporally
resampled to a 100\,ms label grid by average pooling, projected to $d_{\mathrm{proj}}=256$ by separate linear layers, concatenated, and passed through a linear fusion layer. The fused sequence feeds a SELD head with the temporal architecture of the SELDNet~\cite{seldnet}
baseline (bidirectional GRU followed by multi-head self-attention), whose final linear layer produces ACCDOA predictions per 100\,ms frame.

\noindent\textbf{Baselines.}
We compare the performance of \textsc{Bearings} against two baselines: (i) a fully supervised SELD reference trained from
scratch on the 7-channel FOA input (SELDNet, 
results taken from DCASE baselines~\cite{dcase2021, starss23}), (ii) a self-supervised model pre-trained on 7-channel Ambisonics audio (log mel + IVs), evaluated with its encoder frozen under
the same downstream protocol and SELD head (GRAM-Ambisonics~\cite{gram}). Importantly, we compare against the DCASE baseline and GRAM-Ambisonics rather than challenge winning systems, which use ensembling, extensive data augmentation, external synthetic training data, and task-specific architectures trained end-to-end. Our claim concerns the transferability of frozen soundfield embeddings under a fixed lightweight head (1.3–1.9 M trainable parameters), not absolute SELD performance.

\noindent\textbf{Ablation experiments.} To isolate the benefit of the soundfield embeddings produced by \textsc{Bearings} on SELD performance, we compare performance to a single-channel, acoustic pipeline in which the acoustic embeddings produced by the acoustic encoder are fed directly into the SELD head.  To assess the impact of spatial pre-training, we replace the pre-trained soundfield encoder with a small trainable convolutional front-end (two Conv2D layers) that maps the 7-channel FOA input to spatial embeddings, which are fused with the acoustic embeddings.

Furthermore, to measure what spatial information is encoded by the embeddings produced by \textsc{Bearings} independently of any SELD head, we test sound localization performance and reverberation estimation ($RT_{60}$) in simulated spatial scenes. To this end, we render single-source FOA scenes in Matterport3D~\cite{matterport3d} houses with known source
azimuth/elevation and reverberation time. We obtain clip-level soundfield embeddings by mean-pooling the~\textsc{Bearings} embeddings. Each quantity is read out with a $k$-nearest-neighbour regressor fit on training houses and evaluated on held-out houses. As a reference, we apply the same readout to
hand-crafted spatial statistics computed directly from the input features (mean and standard deviation of the active-intensity vectors, short- and long-window diffuseness averages and their difference, and $B_{00}$
log-mel statistics). 

\noindent\textbf{Hand-crafted spatial features.}
As a parameter-free baseline, we assess performance when hand-crafted spatial statistics are fused to the single-channel, acoustic embeddings instead of the soundfield embeddings produced by \textsc{Bearings}. Spatial features consist of the framewise active-intensity vectors of Eq.~(\ref{eq:aiv}), their 50 ms- and 350 ms-smoothed versions, and the magnitudes of the two smoothed vectors. For DOA and reverberation estimation in Matterport3D scenes, each clip is represented by the mean and standard deviation of its AIVs, average short- and long-window diffuseness and their difference, and statistics of the $B_{00}$ log-mel spectrogram.

\noindent\textbf{Metrics.} We report the location-dependent error rate and F-score computed with a
$20^\circ$ spatial threshold ($\mathrm{ER}_{20^\circ}$, $\mathrm{F}_{20^\circ}$) together with the class-dependent
localization error $\mathrm{LE}_{\mathrm{CD}}$ (angular error in degrees; a class with no detection is assigned $180^\circ$) and localization
recall $\mathrm{LR}_{\mathrm{CD}}$~\cite{dcase2021}. The
$95\%$ confidence intervals (Table~\ref{tab:downstream}) are jackknife intervals computed over the test set. For localisation and reverberation estimation on the simulated spatial scenes (Matterport3D probe), we report median angular error and Pearson~$r$ for RT$_{60}$.

\subsection{Implementation details}
During pre-training, we randomly draw two 2\,s crops from each 5\,s clip. We
pre-train models for 50 k steps with AdamW
($\beta_1{=}0.9$, $\beta_2{=}0.98$, peak learning rate
$2\times10^{-4}$, weight decay $0.01$, batch size 256, cosine decay after
10k warm-up steps). For the downstream evaluation, we follow the official DCASE baseline protocol of each dataset~\cite{dcase2021, starss23}. Only the fusion layers and SELD head are trained ($1.9$\,M trainable parameters with GRAM-Ambisonics, $1.3$\,M
with SPEAR and Dasheng), using Adam (learning rate 0.001) with an
MSE loss.

\section{Results}
\label{sec:results}
\subsection{Downstream SELD Performance}
\label{sec:downstream_results}

Table~\ref{tab:downstream} shows that single-channel acoustic encoders retain detection ability ($\mathrm{LR}_{\mathrm{CD}}$ of 35--43 on TAU-NIGENS 2021) but do not localize them: location-dependent F-scores do not exceed 3.2, and angular errors are large. Adding \textsc{Bearings} embeddings to the single-channel encoders yields large gains for each encoder on both datasets without updating either encoder. With GRAM-Clean, the fused model surpasses SELDNet on every metric of both datasets except a tie on STARSS23 $\mathrm{LE}_{\mathrm{CD}}$ ($22.0^\circ$), and on TAU-NIGENS 2021 surpasses GRAM-Ambisonics on every metric. On STARSS23 it improves $\mathrm{ER}_{20^\circ}$ (0.51 vs.\ 0.56), while the intervals for $\mathrm{F}_{20^\circ}$ and $\mathrm{LE}_{\mathrm{CD}}$ contain GRAM-Ambisonics's values and $\mathrm{LR}_{\mathrm{CD}}$ is lower (55.5 vs.\ 58.8). Dasheng and SPEAR follow the same pattern with smaller margins: both surpass SELDNet on every TAU-NIGENS 2021 metric, with Dasheng within 2.7 F-score points of the GRAM-Clean pairing.

\label{sec:label_eff}
\begin{table}[!t]
\centering
\caption{Label efficiency on TAU-NIGENS 2021.}
\label{tab:label_eff}
\setlength{\tabcolsep}{4pt}
\scriptsize
\begin{tabular}{l l cccc}
\toprule
\textbf{Training data} & \textbf{Embeddings}
 & $\mathrm{ER}_{20^\circ}\downarrow$
 & $\mathrm{F}_{20^\circ}\uparrow$
 & $\mathrm{LE_{CD}}\downarrow$ & $\mathrm{LR_{CD}}\uparrow$ \\
\midrule
\multirow{2}{*}{25\%}
 & Bearings (ours)
 & \textbf{0.67} & \textbf{30.6} & \textbf{24.4} & \textbf{49.4} \\
 & Learned-In Domain
 & 0.74 & 21.7 & 40.5 & 45.2 \\
\midrule
\multirow{2}{*}{50\%}
 & Bearings (ours)
 & \textbf{0.57} & \textbf{46.6} & \textbf{20.5} & \textbf{64.2} \\
 & Learned-In Domain
 & 0.66 & 31.7 & 24.8 & 56.5 \\
\midrule
\multirow{2}{*}{100\%}
 & Bearings (ours)
 & \textbf{0.55} & \textbf{50.4} & \textbf{19.3} & \textbf{66.8} \\
 & Learned-In Domain
 & 0.60 & 39.0 & 22.6 & 64.3 \\
\bottomrule
\end{tabular}
\end{table}

\subsection{Ablation Studies}
\label{sec:ablations}

\textbf{Label Efficiency.} Table~\ref{tab:label_eff} shows that our soundfield embeddings consistently outperform the learned embeddings across all data fractions and metrics. Notably, with only 25\% of the training data, our method achieves a relative improvement of 9.5\% in ER20$^{\circ}$ (0.67 vs. 0.74) and 41.0\% in F20$^{\circ}$ (30.6 vs. 21.7) over the baseline, demonstrating superior data efficiency. The advantage persists with 50\% and 100\% data, where our approach maintains higher localization accuracy and recall, as indicated by the improved $LE_{CD}$ and $LR_{CD}$ scores.

\noindent\textbf{Components.} Table~\ref{tab:ablation} (row 2) shows that a randomly initialized, frozen
soundfield encoder substantially underperforms on every metric, confirming that the fusion head reads out structure acquired during pre-training. Acoustic conditioning contributes consistently across metrics (row 3). Removing the direction loss collapses localization ($\mathrm{DOA}$ $49.2^\circ$, row 5) but slightly improves RT$_{60}$ correlation ($r{=}0.55$), suggesting the two objectives trade off. The diffuseness loss adds a modest SELD gain ($\mathrm{F}_{20^\circ}$ $48.3 \rightarrow 50.4$, row 4). Finally, adding hand-crafted spatial features instead of the soundfield embeddings produced by \textsc{Bearings} results in similar performance on localization recall, but the
\textsc{Bearings} soundfield embeddings remain ahead on every other metric (bottom row).

\begin{table}[!t]
\centering
\caption{Ablation studies (TAU-NIGENS 2021, 
Matterport3D).}
\label{tab:ablation}
\setlength{\tabcolsep}{2pt}
\scriptsize
\begin{tabular}{l cccc cc}
\toprule
& \multicolumn{4}{c}{\textbf{SELD}} & \multicolumn{2}{c}{\textbf{Probe}} \\
\cmidrule(lr){2-5}\cmidrule(lr){6-7}
\textbf{Configuration}
 & $\mathrm{ER}_{20^\circ}\downarrow$
 & $\mathrm{F}_{20^\circ}\uparrow$
 & $\mathrm{LE_{CD}}\downarrow$
 & $\mathrm{LR_{CD}}\uparrow$
 & $\mathrm{DOA}^\circ\downarrow$
 & $\mathrm{RT}_{60}$ $r\uparrow$ \\
\midrule
Full model
 & \textbf{0.55} & \textbf{50.4} & \textbf{19.3} & \textbf{66.8}
 & \textbf{16.1} & 0.53 \\
\midrule
Random init
 & 0.70 & 25.8 & 28.2 & 55.4
 & 30.4 & 0.26 \\
No conditioning
 & 0.57 & 46.4 & 20.6 & 65.2
 & 17.5 & 0.50 \\
w/o $\mathcal{L}_\Psi$
 & 0.57 & 48.3 & 19.8 & 65.9
 & 16.1 & 0.52 \\
w/o $\mathcal{L}_q$
 & 0.68 & 29.5 & 27.7 & 63.7
 & 49.2 & \textbf{0.55} \\
 \midrule
 \textit{Hand-crafted} & 0.61 & 43.2 & 22.9 & 66.6 & 23.5 & 0.36 \\
\bottomrule
\end{tabular}%
\end{table}

\begin{table}[!t]
\centering
\caption{Impact of spatial pre-training (TAU-NIGENS 2021).}
\label{tab:ssl_help}
\setlength{\tabcolsep}{3.5pt}
\renewcommand{\arraystretch}{0.95}
\scriptsize
\begin{tabular}{l cccc}
\toprule
\textbf{Encoder initialization, downstream}
 & $\mathrm{ER}_{20^\circ}\downarrow$
 & $\mathrm{F}_{20^\circ}\uparrow$
 & $\mathrm{LE_{CD}}\downarrow$
 & $\mathrm{LR_{CD}}\uparrow$ \\
\midrule
Single-channel, acoustic only       & 0.83 & 2.8  & 102.1 & 34.7 \\
Bearings: random, frozen          & 0.70 & 25.8 & 28.2 & 55.4 \\
Bearings: random, trained end-to-end         & 0.65 & 35.7 & 23.7  & 63.6 \\
Bearings: Pre-trained, frozen     & 0.55 & 50.4 & 19.3  & \textbf{66.8} \\
Bearings: Pre-trained, fine-tuned & \textbf{0.54} & \textbf{51.2} & \textbf{18.1} & 65.5 \\
\bottomrule
\end{tabular}
\end{table}
\noindent\textbf{Impact of self-supervised pre-training.}
Table~\ref{tab:ssl_help} varies the soundfield encoder's initialization
and downstream training setting while keeping the acoustic content
encoder frozen. The pre-trained, frozen \textsc{Bearings} encoder
outperforms its counterpart trained from scratch, increasing
$\mathrm{F}_{20^\circ}$ from $35.7$ to $50.4$ and reducing
$\mathrm{LE}_{\mathrm{CD}}$ from $23.7^\circ$ to $19.3^\circ$.
Under the tested training recipe, hand-crafted spatial features also
outperform the encoder trained from scratch
(Table~\ref{tab:ablation}). Fine-tuning the pre-trained encoder
provides modest further improvements in $\mathrm{F}_{20^\circ}$
($50.4 \to 51.2$) and $\mathrm{LE}_{\mathrm{CD}}$
($19.3^\circ \to 18.1^\circ$), with a small decrease in
$\mathrm{LR}_{\mathrm{CD}}$ ($66.8 \to 65.5$).
These results demonstrate the benefit of spatial pre-training for
downstream SELD and support reusing the learned soundfield embeddings with frozen encoders.

\section{Conclusion}
\label{sec:conclusion}
We introduced \textsc{Bearings}, a self-supervised framework that learns soundfield embeddings from FOA recordings by reconstructing masked direction fields and diffuseness. Attached to three architecturally distinct frozen single-channel encoders through a lightweight fusion head, the embeddings enable sound event localization and detection without retraining either model. These results show that spatial structure can be learned without spatial annotations, with the resulting encoder reused across different frozen acoustic backbones. Future work includes scaling the training corpus and integrating the soundfield embeddings into spatial audio LLMs.

\section{Acknowledgments}
This project received funding from the NWO Talent Program (VI.Veni.202.184; KH). This work used the Dutch national e-infrastructure with the support of the SURF Cooperative using grant no. EINF-14624.

\bibliographystyle{IEEEbib}
\bibliography{strings,refs}

\end{document}